\documentclass[aps,prd,superscriptaddress,preprintnumbers,notitlepage]{revtex4-1}
\usepackage{amsmath}
\usepackage{latexsym}
\def\bal#1\eal{\begin{align}#1\end{align}}

\begin{document}

\preprint{RUP-20-17, RESCEU-9/20}

\title{A new mechanism for freezing extra dimensions with higher-order curvature terms}

\author{Hiroaki W. H. Tahara}
\affiliation{Department of Physics, Rikkyo University, Toshima, Tokyo 171-8501, Japan}

\author{Tsutomu Kobayashi}
\affiliation{Department of Physics, Rikkyo University, Toshima, Tokyo 171-8501, Japan}

\author{Jun'ichi Yokoyama}
\affiliation{Research Center for the Early Universe (RESCEU), Graduate School of Science,
The University of Tokyo, Tokyo 113-0033, Japan}
\affiliation{Department of Physics, Graduate School of Science,
The University of Tokyo, Tokyo 113-0033, Japan}
\affiliation{Kavli Institute for the Physics and Mathematics of the Universe (Kavli IPMU), WPI, UTIAS,
The University of Tokyo, Kashiwa, Chiba 277-8568, Japan}
\affiliation{Trans-scale Quantum Science Institute, The University of Tokyo, Tokyo 113-0033, Japan}

\begin{abstract}
We construct a model of higher dimensional cosmology in which extra dimensions are 
frozen by virtue of the cubic-order Lovelock gravity throughout the cosmic history
from inflation to the present with radiation and matter-dominated regimes in between.
\end{abstract}

\maketitle

\section{Introduction}
	Inflation in the early universe (see e.g.~\cite{Sato:2015dga} for a review) practically realizes isotropic universe even if we start from high degree of anisotropy.
	This is guaranteed by Wald's cosmic no hair theorem of the universe with a positive cosmological constant
	which applies under several conditions such as the strong and dominant energy conditions~\cite{Wald:1983ky}. 
	To realize anisotropic inflation, one must therefore introduce anisotropic matter content breaking 
	the strong energy condition such as a vector field~\cite{Watanabe:2009ct}.
	
	The above is the case of general relativity,
	and the story is totally different in generalized scalar-field theories with nonminimal derivative coupling to the metric 
	which is provided by the generalized Galileon~\cite{Deffayet:2011gz} 
	or the Horndeski theory~\cite{Horndeski:1974wa,Kobayashi:2011nu,Kobayashi:2019hrl}.
	Their higher-order curvature term allows the system to approach a nontrivial attractor 
	and to have anisotropic expansion rates~\cite{Tahara:2018orv}.
	The universe can be spontaneously anisotropized in the presence of such a higher-order curvature term.
	Expansion rates of two spatial dimensions vanish in the limit of large coefficient of the higher-order curvature term.
	The purpose of this letter is to show that such an anisotropic attractor also exists in higher dimensional models
	and it can be used to stabilize the extra dimensions.
	
	Higher dimensions may play an important role in the unification of fundamental forces.
	A pioneering attempt using a higher dimension has been given by Kaluza and Klein~\cite{Kaluza:1984ws, Klein:1926tv}.
	Superstring theory also requires ten-dimensional spacetime to 
	provide a consistent theory of quantum gravity~\cite{Green:1987sp, Green:1987mn}.
	Observationally the dynamics of extra dimensionas would manifest itself in time variation of fundamental constants 
	such as Newton's gravitaional constant and the fine structure constant, which have been severely constrained~\cite{Uzan:2010pm}.
	Hence the extra dimensions must be stabilized in sensible higher-dimensional theories.
	In superstring theories, flux compactification is used for this purpose (\textit{e.g.}~\cite{Douglas:2006es}).

	Here we will propose a different way to \textit{freeze} the extra dimensions with higher-order curvature term 
	in the Lovelock theory~\cite{Lovelock:1971yv}, 
	which is a special case of the Horndeski or generalized Galileon theories.
	The Lovelock theory is the most general gravity theory which consists only of metric degrees of freedom.
	Its evolution equations contain up to the second derivative but it is not always linear.
	In four dimensions, the Lovelock theory is reduced to general relativity with a cosmological constant.
	There have been several studies on the cosmological dynamics in higher dimensions in the Lovelock theory~\cite{MuellerHoissen:1985mm, Deruelle:1989fj, MenaMarugan:1992ig, Maeda:2004hu, 
	Akune:2006dg, Canfora:2008iu, Chirkov:2014nua, Chirkov:2015kja, Kastor:2015sxa, Canfora:2016umq, Chirkov:2019qbe}, 
	and they have not manifestly pointed out the presense of nontrivial attractors.

\section{Anisotropic attractor}	
\subsection{Lovelock theory}
	We start with the Lovelock action up to the cubic order with matter Lagrangian $\mathcal{L}_\textrm{mat}$
	\bal
	&S + S_\mathrm{mat} = \int  d^{(D+1)}x \sqrt{-g} \left( \mathcal{L} + \mathcal{L}_\mathrm{mat} \right) \label{LovelockAction}
	\\
	& { \mathcal{L } \equiv  -\Lambda + \sum_{m=1}^{3} \frac{2m-1}{(2m)!} \kappa_m^{-1} {\mathcal{L}_m} }
	\\
	&\mathcal{L}_m \equiv 2^{-m} \delta^{\mu_1 \mu_2 \ldots \mu_{2m}}_{\nu_1 \nu_2 \ldots \nu_{2m}} \prod_{i=1}^{m} R_{\mu_{2i-1} \mu_{2i}}^{\nu_{2i-1} \nu_{2i}}
	\eal
	where $R_{\mu_{2i-1} \mu_{2i}}^{\nu_{2i-1} \nu_{2i}}$ is the Riemann curvature tensor
	and $\delta^{\mu_1 \cdots \mu_{N} }_{\nu_1 \cdots \nu_{N} }$ 
	is the generalized Kronecker delta. 
	We have defined $\kappa_m$ as constants with mass dimension $2m-D-1$.
	We assume that the space is flat and use the Kasner metric
	\begin{align}
	ds^2= - dt^2  + a_{(i)}^2 \delta_{ij} dx^i dx^j ,
	\label{metricinhigherdim}
	\end{align}
	where $a_{(i)}=a_{(i)}(t)$ is the scale factor in the direction of $x^i$.
	This leads to $\mathcal{L}=-\Lambda-\sum_{m} \kappa_m^{-1} s_{2m}$,
	where $s_{2m}$ is the $2m$-th order symmetric polynomial of $H_{(i)}$'s.
	From Eqs.~\eqref{LovelockAction} to~\eqref{metricinhigherdim}, 
	we obtain the Hamiltonian constraint and evolution equations
	\bal
		&\rho + \Lambda - \sum_{m=1}^{3}  (2m-1)  \kappa_{m}^{-1}  s_{2m}  = 0,
		\label{Hamiltonianeqwithmatter}
	\eal
	\bal
		{p_{(i)}} +  \mathcal{L}
		+ \frac{1}{V} \frac{d}{dt} \left\{ V \sum_{m=1}^{3} \kappa_{m}^{-1} \frac{\partial s_{2m}}{\partial H_{(i)}} \right\} = 0 
		\nonumber \\
		\qquad \textrm{~for~} {i=1,\ldots,D},
		\label{evolutioneqwithmatter}
	\eal
	where 
	$V\equiv\prod_{i} a_{(i)}$ is the volume factor,
	$H_{(i)} \equiv {\dot a_{(i)}}/{a_{(i)}}$ is the expansion rate along $x^i$-axis,
	and $\rho$ and $p_{(i)}$ are defined by
	\bal
	\rho \equiv \frac{2 g_{00}}{\sqrt{-g}} \frac{\delta S_\mathrm{mat}}{\delta g_{00}},
	\quad
	p_{(i)} \equiv \frac{2 g_{ii}}{\sqrt{-g}} \frac{\delta S_\mathrm{mat}}{\delta g_{ii}}.
	\eal
	{In this letter, we consider the cases that the stress-energy tensor is diagonal, 
	which is represented only by $\rho$ and $p_{(i)}$.}
	
	The $(2m)$-th symmetric polynomial $s_{2m}$ is explicitly defined as
	\begin{align}
		&s_2 =  \sum_{1 \le i < j \le D} H_{(i)} H_{(j)} ,
		\\
		&s_4 =  \sum_{1 \le i < j < k < l \le D} H_{(i)} H_{(j)} H_{(k)} H_{(l)} ,
		\\
		&s_6 =  \sum_{1 \le i < j < k < l < m < n \le D} H_{(i)} H_{(j)} H_{(k)} H_{(l)} H_{(m)} H_{(n)}.
	\end{align}
	

\subsection{Attractors under isotropic pressure}
	We will show that there exist attractors 
	if the spacetime is filled with energy contents with isotropic pressure, which means $p_{(i)} = p$ for all $i$.
	Subtracting different components of Eqs.~\eqref{evolutioneqwithmatter}, we yield
	\begin{align}
		\frac{d}{dt} \left\{ V  \left( H_{(j)} - H_{(k)} \right) 
		\sum_{m=1}^{3} \kappa_{m}^{-1} \frac{\partial^2 s_{2m}}{\partial H_{(j)} \partial H_{(k)}} \right\} = 0 
		\nonumber \\
		\textrm{~~~~for $1 \le j < k \le D$} .
		\label{subeq}
	\end{align}
	We integrate these and get their solutions
	\begin{align}
		\left( H_{(j)} - H_{(k)} \right) 
		\sum_{m=1}^{3} \kappa_{m}^{-1}  \frac{\partial^2 s_{2m}}{\partial H_{(j)} \partial H_{(k)}} = \frac{ \mathcal{A}_{jk} }{V}
		\nonumber\\
		\textrm{~~~~for $1 \le j < k \le D$} ,
		\label{subeq2}
	\end{align}
	where $\mathcal{A}_{jk}'s$ are integration constants.

	In principle, full set of Eqs.~\eqref{Hamiltonianeqwithmatter} and \eqref{evolutioneqwithmatter} 
	describes evolution of the system in the phase space $\{ a_{(i)} , H_{(i)} \}$,
	but we try to see what the subset \eqref{subeq} or \eqref{subeq2} indicates.
	The left-hand-side of Eq.~\eqref{subeq2} is a function of only $H_{(i)}$'s, and if we know how $V$ evolves, 
	we can track the evolution of the system in a phase space $\{ H_{(i)} \}$.
	When $V$ is an increasing function of time,
	the system approaches the regions where the following equations are satisfied.
	\begin{align}
		\left( H_{(j)} - H_{(k)} \right) & \sum_{m=1}^{3} \kappa_{m}^{-1} \frac{\partial^2 s_{2m}}{\partial H_{(j)} \partial H_{(k)}} = 0 .
		\label{fixedpointeq}
	\end{align}
	A similar equation has been derived in \cite{Chirkov:2015kja} under the assumption that all  $H_{(i)}$'s are constant. 
	In deriving Eq.~\eqref{fixedpointeq}, however, we do not need to restrict $H_{(i)}$'s to be constant, so that we can trace the entire evolution
	of the universe as we will see below.
	We name those regions ``{(an)isotropic attractors}'',
	although we must carefully analyze the system to see whether they really act as an attractor.

\subsection{Classification of attractors}
	Here we classify the roots of Eq.~\eqref{fixedpointeq} into several types.
	On these roots, some of $H_{(i)}$'s have the same value, and thus we label the attractors with ${N}_\textrm{diff}$,
	which denotes the number of different values of $H_{(i)}$'s.

\paragraph{Isotropic case $(N_{\rm diff}=1)$}
	When all of $H_{(i)}$'s are the same, all of the equations \eqref{fixedpointeq} are trivially satisfied.
	The universe exhibits isotropic expansion and we just call it isotropic attractor.

\paragraph{Anisotropic case $(N_{\rm diff} \ge 2)$}
	For the simplest departure from isotropy, we consider the case
	in which $H_{(i)}$'s take two different values, $\alpha$ and $\beta$.
	\begin{align}
		H_{(i)} &= \alpha \textrm{~~~~for~} 1\le i \le d,  \label{subalpha}
		\\
		H_{(i)} &= \beta \textrm{~~~~for~}  d+1 \le i \le D ,  \label{subbeta}
	\end{align}
	where $d$ is any integer satisfying $2 \le d \le D-1$.
	This corresponds to the case for $N_{\rm diff} = 2$.
	Then Eq.~\eqref{fixedpointeq} gives the relation between $\alpha$ and $\beta$
	\begin{align}
		\sum_{m=1}^{3} \kappa_{m}^{-1}  {Q}_{m}(\alpha,\beta) = 0 ,
		\label{alphabeta}
	\end{align}
	where
	\begin{align}
		{Q}_{m}(\alpha,\beta) &\equiv \left. \frac{\partial^2 s_{2m}}{\partial H_{(j)} \partial H_{(k)}} \right|_{H_{(1\le j \le d)}=\alpha, H_{(d+1 \le k \le D)}=\beta } 
		\nonumber \\
		&= \sum_{l=0}^{2m-2} \left(\begin{array}{@{}c@{}} d-1 \\ l \end{array} \right)
		\left(\begin{array}{@{}c@{}} D-d-1 \\ 2m-2-l \end{array} \right) \alpha^l \beta^{2m-2-l}.
	\end{align}

	The discussion above can be generalized straightforwardly to the cases with larger $N_\textrm{diff}<D$.
	In this letter, we focus on the case for $N_\textrm{diff}=2$ for simplicity.

\section{Large universe and frozen extra dimensions} \label{secFreezing}
\subsection{Evolution scenario}
	In order to compactify the extra dimensions successfully, 
	we have to explain both why they are small and stable.
	Otherwise, we could observe Kaluza-Klein particle in a particle accelerator or varying Newton's constant~\cite{Uzan:2010pm}.
	We try to explain those properties by realizing $\beta \ll \alpha$ in the whole cosmic history from inflation, 
	where we regard $\alpha$ and $\beta$ as the expansion rate of $(d+1)$-dimensional universe and the extra $(D-d)$ dimensions.

	We consider the	standard history of inflationary cosmology.
	That is, we set the initial condition at the beginning of inflation, 
	followed by reheating regime dominated by coherent field oscillation of the inflaton
	in the case of standard potential-driven models or by kinetic energy in the case of 
	k- or G-inflation models~\cite{ArmendarizPicon:1999rj, Kobayashi:2010cm}.
	Then the universe turns to be radiation dominant with a reheating temperature $T_R$
	to trace the standard thermal history of the universe in Big Bang cosmology toward matter domination
	and dark energy domination.
	
	We set the topology of space as $D$-dimensional tori and assume that the size or period of each dimension 
	is of the same order of magnitude given by $\ell$, simply from democratic viewpoint, 
	at the beginning of inflation of $d$-dimensional space with the Hubble parameter $\alpha$.
	
	We assume that inflation and subsequent reheating are realized by an appropriate scalar-field model 
	which has a practically homogeneous configuration over $T^D$ initially and 
	anisotropic inflation is realized in a similar manner as in the previous paper~\cite{Tahara:2018orv}.
	This initial condition is no less natural than required in conventional three dimensional inflation models 
	if $\ell$ is of the order of $\alpha^{-1}$ or smaller.
	With $\beta \ll \alpha$ for a sufficiently long time $\tau \gg \alpha^{-1}$,
	the $d$-dimensional space becomes exponentially large while
	freezing the extra dimensions practically.
	
	After inflation the universe will be eventually dominated by radiation through reheating process
	intrinsic to each inflation model.
	For simplicity, we consider the case that reheating temperature $T_R$ is much smaller than 
	the inverse size of the extra dimension, $\ell^{-1}$, so that equation-of-state parameter $w$ is
	anisotropic in this regime, namely, $p_{(i)}/\rho=1/d$ in large $d$ dimensions, 
	and $p_{(i)}=0$ in the small extra dimensions.
	As the universe becomes matter ($w=0$) or dark-energy ($w\approx -1$) dominant, 
	the equation of state becomes isotropic again.
		
	We will concentrate on the case for ten-dimensional spacetime with $D=9$ and $d=3$, since it is of our most interest.
	First, we consider condition for freezing extra dimensions $\beta \ll \alpha$ when pressure is isotropic.
	We need other treatment in the radiation-dominated era and we will find that the radiation also makes
	expansion rate of the extra dimensions be strongly suppressed.

\subsection{Under isotropic pressure} \label{subsecIsotropic}
	First we consider an anisotropic attractor under isotropic pressure.
	The anisotropic attractor is allowed to exist when the pressure is isotropic.
	Let us investigate the roots of Eq.~\eqref{alphabeta} for $\beta$
	and show some of them can satisfy $\beta \ll \alpha$.
	When the root $\beta=\beta(\alpha)$ is much smaller than $\alpha$, 
	$\beta(\alpha)$ is given by solving
	\bal
	\kappa_1^{-1} + \kappa_2^{-1} \alpha^2 + 10 \kappa_3^{-1} \alpha^2 \beta^2 = 0,
	\label{alphabetaAsymptotic}
	\eal
	which we get by neglecting higher-order terms of $\beta$ from Eq.~\eqref{alphabeta}.
	We immediately obtain
	\bal
	\beta(\alpha) = \pm \sqrt{ - \frac{ \kappa_1^{-1} + \kappa_2^{-1} \alpha^2 }{ 10 \kappa_3^{-1} \alpha^2 } },
	\label{BetaOfAlpha}
	\eal
	where $\kappa_3 < 0$ is assumed. 
	
	We now define lower bound of $\alpha$ as $\alpha_\textrm{min}$, 
	since our Universe seems to expand exponentially today.
	We require $\beta(\alpha) \ll \alpha$ to be satisfied for any $\alpha \ge \alpha_\textrm{min}$ 
	for consistency.
	From Eq.~\eqref{BetaOfAlpha}, we get
	\bal
	- \kappa_3/\kappa_1 \ll \alpha_\textrm{min}^4 , \qquad - \kappa_3/\kappa_2 \ll \alpha_\textrm{min}^2.
	\label{ConsistencyCondition}
	\eal

	Substituting $\beta=\beta(\alpha)$ into the Hamiltonian constraint~\eqref{Hamiltonianeqwithmatter},
	we get the effective constraint at the leading order
	\bal
	{\rho} + \Lambda - 3 \kappa_{1}^{-1} \alpha^2 
	- 8 \kappa_{2}^{-1} \alpha^3 \beta(\alpha) = 0. \label{EffectiveConstraint}
	\eal
	This corresponds to the Friedmann equation with an additional term.
	It brings us a new effect, which may matter in the very early universe unless $\kappa_2^{-1} = 0$.
	When $\kappa_1^{-1}$ and $\kappa_2^{-1} \alpha \beta(\alpha)$ are comparable and $\rho + \Lambda$ is negligible,
	the universe expands with almost constant rate
	\bal
	\alpha = \mp \sqrt{ - \frac{ 45 \kappa_2^{3} }{ 32 \kappa_1^{2} \kappa_3 } } \times \mathrm{sign}(\kappa_1 \kappa_2).
	\eal
	
	If we can neglect the additional term for any $\alpha$ of interest, the observed universe always obeys 
	the same Friedmann equation as in general relativity.
	In this case, since $\beta \ll \alpha$,
	only $\alpha$ affects the evolution of density of energy contents and nothing feels extra dimensions.
	
	In the limit $\kappa_ 3\to 0$, we have $\beta(\alpha) \to 0$ for fixed $\alpha$.
	This means that it is possible to realize arbitrarily small $\beta$.
	In the limit $\alpha \to \infty$, $\beta(\alpha)$ converges to a finite value,
	and thus we do not suffer from divergent behavior for larger $\alpha$.
	Therefore, we can slow down the extra dimensions arbitrarily even during inflation, 
	as long as isotropic pressure dominates the higher-dimensional spacetime.

\subsection{Radiation-dominated universe} \label{sectionExtraFreezeAnisotropic}
	Next, we consider a radiation-dominated universe,
	where we no longer use the roots \eqref{BetaOfAlpha}.
	We continue using $\alpha$ and $\beta$ instead of $H_{(i)}$, 
	since the expantion rates tend to take two values as in Eqs.~\eqref{subalpha} and \eqref{subbeta} 
	even after entering the radiation-dominated era.
	
	For $D=9$ and $d=3$, if $\kappa_2^{-1}=0$ then radiation allows $\beta=\dot \beta=0$ to be a solution, that is,
	Eqs.~\eqref{Hamiltonianeqwithmatter} and \eqref{evolutioneqwithmatter} yield
	\bal
	\dot \beta(\alpha,\beta)|_{\beta=0} = 0 , 
	\qquad
	\left. \frac{\partial }{\partial \beta} \dot \beta(\alpha,\beta) \right|_{\beta=0} = -3\alpha .
	\eal
	The minus sign of the first derivative shows this solution is stable as long as the universe is expanding.
	
	If $\kappa_2^{-1} \ne 0$, the solution shifts like ($\beta=\beta_0(\alpha)$, $ \dot \beta=0$).
	In general, it is difficult to find an analytic form of $\beta_0$, 
	but its approximate value can be estimated as
	\bal
	\beta_0(\alpha) &\approx  -    \left.{\dot \beta(\alpha,\beta)}/{\tfrac{\partial }{\partial \beta} \dot \beta(\alpha,\beta) }
	\right|_{\beta=0}          
	\nonumber\\
	&= - \frac{\kappa_1^{-1}\kappa_2^{-1} \alpha^3 
	(4\kappa_1^{-2} + 3\kappa_1^{-1}\kappa_2^{-1}\alpha^2+9\kappa_2^{-2}\alpha^4) }
	{ 16\kappa_1^{-4} + 56\kappa_1^{-3}\kappa_2^{-1}\alpha^2 - 9\kappa_1^{-2}\kappa_2^{-2}\alpha^4 
	+ 18 \kappa_1^{-1}\kappa_2^{-3}\alpha^6  + 27 \kappa_2^{-4}\alpha^8 
	+ 20 \kappa_1^{-2}\kappa_2^{-1}\kappa_3^{-1} \alpha^6}.
	\eal
	This expression typically leads to $\beta_0(\alpha) = \mathcal{O} (\beta(\alpha)^2/\alpha)$, 
	where $\beta(\alpha)$ is the function defined in Eq.~\eqref{BetaOfAlpha}.
	This means that in the radiation-dominated era, the expansion rate of the extra dimensions is suppressed more strongly  
	than in the isotropic-pressure-dominated case with the same value of $\rho$.
	Taking this suppression into account, 
	we now succeed in freezing the extra dimensions for the whole standard history.

\section{Discussion}\label{secDiscussion}
	
	As we have seen in the case under isotroic pressure,
	small $\kappa_3$ plays a key role to freeze extra dimensions.
	On the other hand, nonvanishing $\kappa_2^{-1}$ yields the additional term in Eq.~\eqref{EffectiveConstraint}.
	It can cause exponential expansion in the very early universe 
	where the propagation speed of gravitational waves can be varied.
	This is worth considering but it is beyond the scope of this letter
	and we avoid this exponential expansion by taking $\kappa_2 \gg \kappa_1 \alpha_\textrm{max}^2$ 
	in the following discussion,
	where $\alpha_\textrm{max}$ is the maximum of $\alpha$ of interest.

	The $d$-dimensional effective gravitational constant depends on $\beta$ as $\dot G/G = (d-D) \beta$.
	From the constraint $|\dot G/G| = (4\pm 9) \times 10^{-13} ~\textrm{yr}^{-1} $ given by the Lunar laser ranging experiment~\cite{Williams:2004qba} implies 
	that we have to require that in the late universe
	\bal
	| \beta | \lesssim 10^{-13} ~\textrm{yr}^{-1} \sim 10^{-3} H_0,
	\eal
	where $H_0$ is the present Hubble constant.
	From Eq.~\eqref{BetaOfAlpha}, we set $\alpha=H_0$ and get 
	\bal
	- \kappa_3/\kappa_1 \lesssim 10^{-{5}} H_0^4.
	\eal
	Note that although the effective gravitational constant in the Friedmann equation is
	different in principle from Newton's gravitational constant in the Poisson equation of the gravitational potential 
	which is constrained in~\cite{Williams:2004qba}, 
	those constants are approximately the same if the extra dimensions are frozen $\beta \ll \alpha$ 
	and if their size is much smaller than the scale of observation.

	When $\kappa_3$ is small, one might think that such a large higher-order term could 
	decrease the energy scale of the unitarity bound of the model.
	To show that it is not the case around the anisotropic attractor,
	we give a simple estimation of interaction of gravitons $h_{ij}$ 
 on the unperturbed metric $\bar{g}_{\mu\nu}$ which is equal to the metric in Eq.~\eqref{metricinhigherdim}.
	We compare the following coefficients of the perturbed Riemann tensor $\delta {R}_{\mu_1\mu_2}^{ \nu_1\nu_2}$ 
	in the Lagrangian
	\bal
	\kappa_1^{-1} \delta^{\mu_1\mu_2}_{\nu_1\nu_2} 
	, \qquad 
	\kappa_3^{-1} \delta^{\mu_1\mu_2 \cdots \mu_6}_{ \nu_1\nu_2 \cdots \nu_6} \bar{R}_{\mu_3\mu_4}^{ \nu_3\nu_4} \bar{R}_{\mu_5\mu_6}^{ \nu_5\nu_6},
	\eal
	where $\bar{R}_{\mu_1\mu_2}^{ \nu_1\nu_2}$ is the unperturbed Riemann tensor.
	The former are just of the order of $\kappa_1^{-1}$.
	Since $h_{ij}$ appears in $\delta R_{i_1 i_2}^{j_1 j_2}$ and $\delta R_{0i}^{0j}$,
	$\kappa_3^{\!-1} \! \delta^{0 i i_1 i_2 M_1 M_2}_{0 j  j_1j_2 N_1 N_2} \bar{R}_{i_1 i_2}^{j_1 j_2} \bar{R}_{M_1 M_2}^{ N_1 N_2}$ and 
	$\kappa_3^{\!-1} \! \delta^{0 i i_1 i_2 M_1 M_2}_{0 j  j_1j_2 N_1 N_2} \bar{R}_{0 i}^{0 j} \bar{R}_{M_1 M_2}^{ N_1 N_2}$ are largest of all possibilities, where $M,N,\ldots$ denote the indices for the extra dimensions.
	These are of the order of $\kappa_3^{-1} \alpha^2 \beta^2$, which is of the same order of $\kappa_1^{-1} $ 
	as long as $\kappa_2 \gg \kappa_1 \alpha^2$.
	Therefore the self-couplings of gravitons is not so larger than in the general relativity.
	Perturbation theory does not break down until around the Planck scale.
	We have also checked that propagation speed of gravitons is almost the same as speed of light, 
	and we can avoid inconsistency with observations by taking small $\kappa_3$.
	
	Small extra dimensions let 
	the Kaluza-Klein modes obtain large masses
	in both our compactification and the flux compactification.
	In the latter case, the zero modes called moduli also obtain large mass,
	while in our case they are still massless.
	In our calculation, we notice that the coefficient of their kinetic term is much larger 
	if inequalities~\eqref{ConsistencyCondition} are satisfied.
	This fact leads to their strongly supressed coupling to matters via canonicalization of variables.
	For this reasons, we expect that small coupling constant $\kappa_3$ suppresses production of 
	those massless degrees of freedom and astrophysical and cosmological observation 
	is consistent with prediction of general relativity.

\section{Conclusion}\label{secConclusion}
	We have investigated the spatially flat homogeneous model in the Lovelock theory up to the cubic order.
	Under isotropic pressure, 
	subtractions of the evolution equations yield the conservation law \eqref{subeq}.
	It implies that as the total volume expands, the system converge on the roots of Eq.~\eqref{fixedpointeq}, 
	which we call (an)isotropic attractors.
	We have focused especially on the anisotropic attractor with two different expansion rates.
	
	We have considered the possibility to apply the anisotropic attractor to freeze extra dimensions.
	According to the standard cosmology, the universe is dominated by two kinds of energy contents.
	One is energy contents having isotropic pressure, e.g., 
	potential and kinetic energy of a scalar field, cold matter, and the cosmological constant.
	In this isotropic case, we can use the ansiotropic attractor to suppress growth of the extra dimensions 
	compared to the lower-dimensional universe 
	if inequalities~\eqref{ConsistencyCondition} are satisfied.
	The other has anisotropic pressure, such as radiation of sufficiently low energy scale.
	Even though we cannot use the anisotropic attractor, radiation allows the expansion rate of extra dimensions 
	to become smaller 	than in the isotropic case.
	Therefore, we conclude that in the whole of standard cosmic history, 
	extra dimensions are frozen by using the Lovelock theory.
	
	We can straightforwardly extend this mechanism to the Lovelock theory 
	or generalized Galileon up to an arbitrary higher order.
	In the generalized Galileon, odd-power terms of expansion rate appears in the Hamiltonian constraint,
	whereas the Lovelock theory contains only even-power terms.
	The details will be studied in our forthcoming paper.

\begin{acknowledgments}
We thank Taizan Watari and Hideyuki Tagoshi for useful comments.
HWHT was supported by the Grant-in-Aid for JSPS Research Fellow and the Advanced Leading Graduate Course for Photon Science.
TK was partially supported by JSPS KAKENHI Grant Nos.~JP20H04745 and JP20K03936.  
JY was partially supported by JSPS KAKENHI Grant Nos.~JP15H02082, JP20H00151, and Grant-in-Aid for Scientific Research on Innovative Areas JP20H05248.
\end{acknowledgments}
 
\bibliography{mypaper}

\end{document}